\documentclass[reprint,amsmath,amssymb,aps,prd,floatfix]{revtex4-2}

\usepackage{graphicx} 
\usepackage{xcolor} 
\usepackage[colorlinks=true,linkcolor=blue,citecolor=blue,urlcolor=blue]{hyperref} 
\usepackage{enumitem} 

\setlist[enumerate,1]{
    label={\!\!\alph*)},
    leftmargin=*,
    nosep
}

\usepackage[left=1.35cm, 
            right=1.23cm, 
            top=1.73cm, 
            bottom=1.73cm]{geometry}
\usepackage{multirow}  

\graphicspath{{./Figures/}}
\renewcommand{\rm}[1]{\mathrm{#1}}
\newcommand{\dm}[1]{\mathrm{~#1}}

\begin{document}

    \title{Leptonic and hadronic Models of high-energy nebula around V4641 Sgr}

\author{
    Maksim~Kleimenov$^{1,2}$,
    Andrii~Neronov$^{2,3}$,
    Foteini~Oikonomou$^{4}$,
    and Dmitri Semikoz$^{2}$.
}
\email{maksim.kleimenov@polytechnique.edu, andrii.neronov@apc.in2p3.fr, foteini.oikonomou@ntnu.no, dmitri.semikoz@apc.univ-paris7.fr}

\affiliation{
    \mbox{$^1$Institut Polytechnique de Paris, Palaiseau, France}\\
    \mbox{$^2$Université de Paris Cite, CNRS, Astroparticule et Cosmologie, Paris, France}\\
    \mbox{$^3$Laboratory of Astrophysics, Ecole Polytechnique Federale de Lausanne,  Lausanne, Switzerland}\\
    $^4$Institutt for fysikk, NTNU, Trondheim, Norway
}

\date{December 15, 2025}

\begin{abstract}
    A prominent, 200-pc-scale high-energy nebula surrounding the microquasar V4641~Sgr is the brightest known gamma-ray source in the southern sky at $E > 100\dm{TeV}$. 
    In this paper, we develop self-consistent leptonic, hadronic, and leptohadronic models that reproduce both the observed spectrum and morphology of the source. Purely leptonic models are energetically more favorable yet they require rather specific morphological assumptions. The gamma-ray morphology of the source can be better explained within a hadronic scenario based on the identification of cold gas structures spatially correlated with the observed gamma-ray emission. However, a purely hadronic model for the source emission requires a substantial energy reservoir in protons and fails to reproduce the extended x-ray emission recently detected by XRISM. We show that emission including a combination of leptonic and hadronic components can reproduce both the spectral and morphological properties of the source. We provide predictions for the x-ray and neutrino spectra of~the~nebula that can discriminate the hadronic and leptonic contributions to the overall source signal.
\end{abstract}

    \maketitle
    \section{Introduction}

Cosmic rays (CRs) are charged particles with energies up to~\mbox{$10^{20}\dm{eV}$} which come to Earth from space and are locally measured directly with satellites at energies \mbox{$E<100\dm{TeV}$} and indirectly with Earth-based cosmic ray detectors at higher energies. Knowledge of the CR spectrum on Earth is insufficient to infer the distribution of CR sources in the Galaxy. Instead, this distribution can potentially be deduced from $\gamma$-ray and neutrino data. Measurements of diffuse $\gamma$-ray flux are performed by the Fermi-LAT telescope at $E\lesssim 1\dm{TeV}$~\cite{Fermi-LAT:2012edv,Neronov:2019ncc} and primarily by HAWC and LHAASO at $E>1\dm{TeV}$~\cite{HAWC:2023wdq,LHAASO:2023gne,LHAASO:2024lnz}. The contribution of unresolved leptonic sources to the diffuse $\gamma$-ray flux has been shown to be subdominant~\cite{Kaci:2024lwx}, while the hadronic part of the diffuse flux is constrained by the diffuse Galactic neutrino flux observations~\cite{ANTARES:2022izu,IceCube:2023ame,Allakhverdyan2025}.

Comparison of local CR flux measurements and global models of the CR distribution in the Galaxy based on $\gamma$-ray and neutrino data allows us to constrain the properties of the population of cosmic-ray sources responsible for the knee of the CR spectrum~\cite{Prevotat:2024ynu,Castro:2025wgf,Prevotat:2025ktr}. Such sources are possibly rare, so they may not produce continuous diffuse emission~\cite{Giacinti:2023upw} and may instead provide an intermittent, time-variable, kneelike feature in the cosmic-ray spectrum. An example of such rare and powerful sources able to accelerate CRs to the knee energies are microquasars recently detected in the energy range \mbox{$E>100\dm{TeV}$} by LHAASO~\cite{LHAASO2024}. 

Gamma-ray and neutrino data can also be used to trace episodes of PeV cosmic-ray injection by such sources. CRs escaping from the source following an injection episode propagate in the Galactic magnetic field~(GMF), which includes ordered and turbulent components. Such magnetic field structures lead to anisotropic spread of CRs around their sources, with the distribution elongated along the ordered component of the GMF~\cite{Semikoz-MF}. Anisotropic spread of CRs can be more readily traced in the regions outside the Galactic disk, where the field structure is less perturbed by multiple supernovae~\cite{Jansson:2012pc,Unger:2023lob,Korochkin:2024yit}.   

With total flux at $100\dm{TeV}$ exceeding $4\dm{Crab}$, V4641~Sgr, an x-ray binary located $\sim8^\circ$ from the Galactic Center, is the brightest source yet revealed in the highest-energy sky. First detected by HAWC~\cite{HAWC2024}, it was later observed by H.E.S.S. \cite{HESS-AA2025, HESS-2024} and LHAASO~\cite{LHAASO2024}. The extended emission around the source has also been detected by XRISM in the x-ray domain~\cite{Suzuki2025}. Several models of high-energy activity of the source have been considered, including  leptonic~\cite{leptonic2025, Dmytriiev2025},  hadronic~\cite{Neronov2024}, and leptohadronic models~\cite{leptohadronic2025}. Each of these scenarios meets certain difficulties. Leptonic models have problems in explaining  the highest-energy photons far away from the source. To the contrary, hadronic models face difficulties with the overall power budget of the nebula. 

In the present work, we revise leptonic, hadronic, and leptohadronic models, accounting both for spectral and morphological properties of the extended source. In Section~\ref{section:mq}, we summarize recent observation results for V4641~Sgr; in Section~\ref{section:propagation}, we explain the high-energy particle propagation mechanisms in the GMF; Section~\ref{section:leptonic} is dedicated to a discussion of the leptonic model, its features and possible explanations of the nebula's morphology; in Section~\ref{section:leptohadronic}, we describe hadronic and leptohadronic models of the source; we discuss and summarize our results in Section~\ref{section:discussion}.

\section{Microquasar V4641 Sgr}
\label{section:mq}

\subsection{The binary system}

Located at Galactic coordinates \mbox{$l = 6.8^\circ$, $b = -4.8^\circ$}, V4641~Sgr is an x-ray binary consisting of a $6.4\,M_\odot$ black hole and a companion star with a mass of~$2.9\,M_\odot$, a radius of~$5.3\,R_\odot$, and spectral type~B9\,III \mbox{($T_\rm{eff} = 10\,500\dm{K}$)}. The system has an orbital period of 2.8 days and a semimajor axis of \mbox{$a = 17.5\,R_\odot$~\cite{MacDonald2014}}.

An initial lower limit on the binary’s distance from Earth, $d > 0.5\dm{kpc}$, was inferred from radio measurements~\cite{Hjellming2000}. Subsequent optical observations in quiescence yielded \mbox{$d = 6.2 \pm 0.7\dm{kpc}$~\cite{MacDonald2014}}, while parallax measurements gave \mbox{$d = 6.6^{+2.5}_{-1.4}\dm{kpc}$} (\textit{Gaia}~DR2,~\cite{GaiaDR2}) and \mbox{$d = 5.9^{+1.1}_{-0.8}\dm{kpc}$} (\textit{Gaia}~DR3,~\cite{GaiaDR3}). For our analysis, we adopt the weighted average of these measurements: \mbox{$d = 6.1 \pm 0.5\dm{kpc}$}. At this distance, V4641~Sgr is inevitably located close to the Galactic bar~\cite{BlandHawthorn2016}.

\begin{figure}[b]
    \vspace{-2ex}
    \centering
    \includegraphics[width=\linewidth]{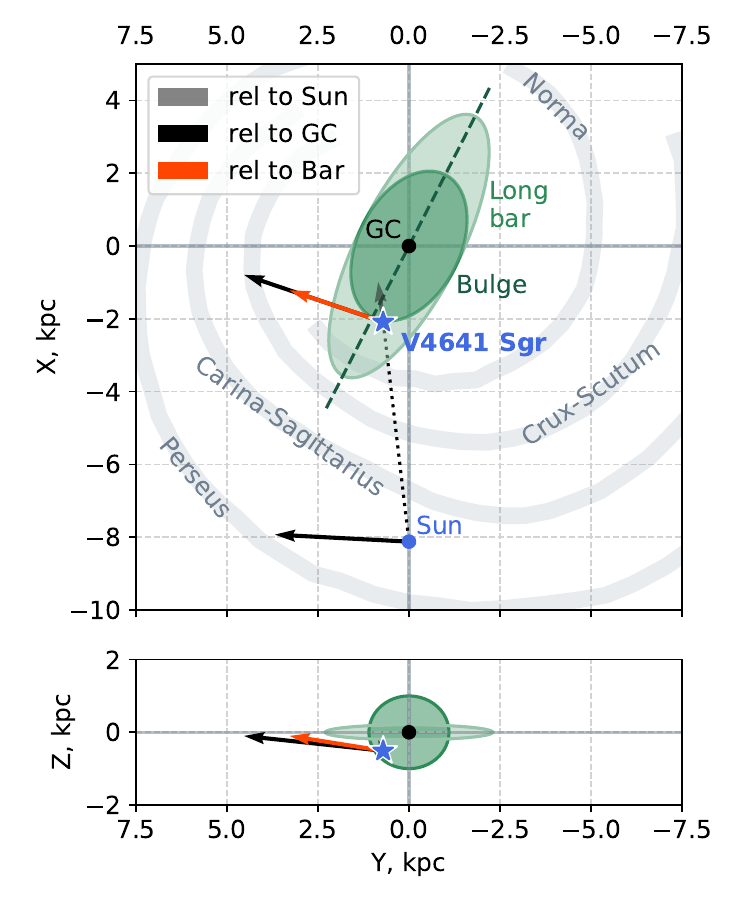}
    \vspace{-5ex}
    \caption{
    Position of V4641~Sgr in the Milky Way relative to the Galactic bar, which consists of the Galactic Bulge (rich green) and the Long bar (pale green) \cite{BlandHawthorn2016}. The upper panel shows the top view in Galactic coordinates, and the lower panel presents the side view. Velocity vectors are shown for the Sun and for V4641~Sgr: the black vectors correspond to velocities relative to the Galactic center, the gray vector shows the motion of V4641~Sgr relative to the Sun~\cite{GaiaDR3,Lindstrom2005}, and the red vector represents the binary’s velocity in the rotating frame of the bar. The approximate positions of the spiral arms are indicated in pale gray \cite{Drimmel2001}.
    }
    \label{fig:source_position}
\end{figure}

Gaia astrometry~\cite{GaiaDR3} and spectroscopic measurements of the binary~\cite{Lindstrom2005} provide the velocity of the source shown in \autoref{fig:source_position}. In this figure, we display the velocity components relative to the Galactic Center and relative to the bar, assuming an angular velocity of \mbox{$\Omega_b = 40\dm{km/s/kpc}$} for the bar~\cite{BlandHawthorn2016}. The binary has a high enough linear velocity in the Galactic bar rest frame, to indicate that the system likely entered the region $\sim 10\dm{Myr}$ ago,
is currently passing through it, and might originate from one of the star-forming regions in the disk outside the bar. A calculation of the binary trajectory in an axisymmetric gravitational potential (without the Galactic bar) shows that the last Galactic plane crossing happened $10\dm{Myr}$ ago~\cite{Salvesen2020}.

V4641~Sgr is famous for its fast and violent outburst in~1999 observed from radio to x-rays. This event included three bright x-ray flares up to 12~Crab \cite{Revnivtsev2002} and~a~superluminal jet \cite{Hjellming2000} with an~inclination angle $\alpha_{\mathrm{jet}} < 16^\circ$~\cite{Salvesen2020}, earning the system the status of~a~''microblazar''~\cite{Chaty2003}.

The 1999 flare was followed by several minor flashes with decreasing magnitude (see Appendix~\ref{appendix:binary_lightcurve}). Spectroscopic studies of major flashes in 1999, 2002, and 2004 report the detection of cold wind outflows from the binary~\cite{Munoz-Darias2018}.

\subsection{Extended gamma-ray and x-ray emission}

\begin{figure}[b]
    \centering
    \includegraphics[width=\linewidth]{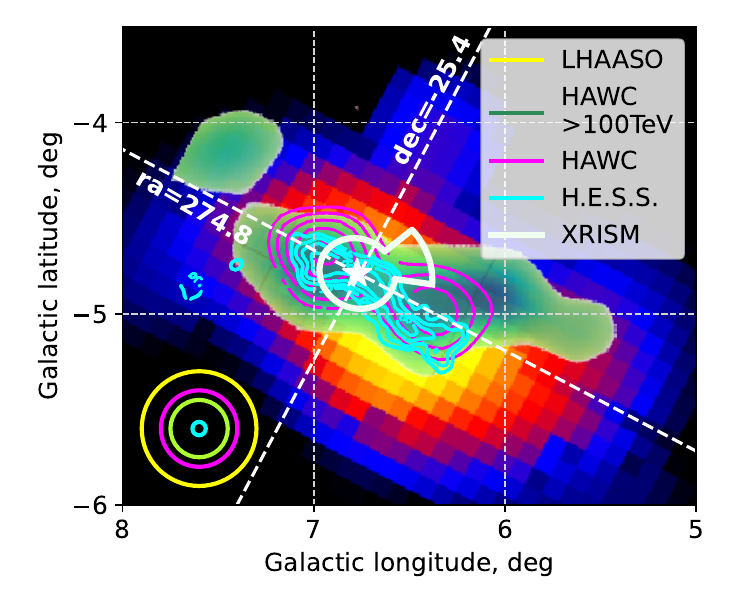}
    \caption{Combined image of H.E.S.S. significance contours~\cite{HESS-2024} (cyan), HAWC contours (magenta) and HAWC~\mbox{$>100\dm{TeV}$} TS~map (blue-green)~\cite{HAWC2024}, and LHAASO TS~map (blue-red-yellow)~\cite{LHAASO2024}. The circles on the bottom left correspond to the point-spread-function (PSF) size of the observations~\cite{HESS-AA2025, HAWC_PSF_Crab}. The XRISM x-ray measurement region boundaries are shown in white~\cite{Suzuki2025}.}
    \label{fig:combined_image}
\end{figure}

Recently, the HAWC, LHAASO, and H.E.S.S. Collaborations have reported the detection of extended high-energy radiation from the source~\cite{HAWC2024, LHAASO2024, HESS-AA2025}. The luminosity of the source at $100\dm{TeV}$ 
reaches 4~Crab units~\cite{LHAASO2024}. A summary of imaging data on the source is shown in~\autoref{fig:combined_image}. The significance maps were transformed into Galactic coordinates with the use of \texttt{astropy}~\cite{Astropy_2022}.

Following the gamma-ray observations, XRISM has found extended x-ray emission in the range from \mbox{$2$ to $10\dm{keV}$} with a width of $\sim10\dm{pc}$ around the source, and power-law index~\mbox{$\sim2$}~\cite{Suzuki2025}. The XRISM \textit{keyholelike} measurement region is also presented in \autoref{fig:combined_image}.

We  notice that the high-energy emission exhibits a filamentary morphology with the following features:
\begin{enumerate}
    \item its size increases with energy;
    \item its shape becomes increasingly asymmetric at higher energies;
    \item it has a nonzero width, resolved by H.E.S.S.~\cite{HESS-AA2025}.
\end{enumerate}

We note also that the major axis of the high-energy nebula is in significant misalignment of $\sim60^\circ$ with the low-energy jet detected in VLBI observations from 1999 and 2004 \cite{Hjellming2000, 2011TwoBHCandidates}. This is, however, questioned by a recent data reanalysis, which reveals a jet almost aligned with the high-energy nebula major axis~\cite{Marti2025}.

Another extended feature, a cloud of $\sim35\dm{pc}$ scale slightly offset from the source, has been detected by MeerKAT in a radio frequency window centered at $1.217\dm{GHz}$~\cite{Grollimund2026}. The clouds's elongation is approximately orthogonal to the gamma-ray nebula axis and forms an angle of $\approx 30^\circ$ with the projected 1999 radio-jet axis~\cite{Hjellming2000}.

\section{Particle Propagation along Galactic Magnetic Field Line}
\label{section:propagation}

We assume that the source morphology is explained by high-energy particle propagation along the direction of the GMF, similar to Ref.~\cite{Neronov2024}. The magnetic field itself consists of regular and turbulent components $B_0$ and $\delta B$. Their ratio \mbox{$\eta = \delta B / B_0$} defines the propagation regime, which varies from free streaming of the particles for \mbox{$\eta \ll 1$,}~anisotropic diffusion for $\eta\sim1$, and isotropic diffusion for \mbox{$\eta\gg 1$ \cite{SemikozRev2019}}.

Although the GMF direction in the vicinity of the source is unknown,
the GMF in the Milky Way halo consists of toroidal and ``x-shaped'' components~\cite{Jansson:2012pc,Unger:2023lob,Korochkin:2024yit}, which at the location of the source are responsible for the horizontal and vertical components of the magnetic field relative to the Galactic plane, respectively. The combination of these components must result in the observed direction of elongation of the nebula.

\begin{figure}[b]
    \centering
    \includegraphics[width=\linewidth]{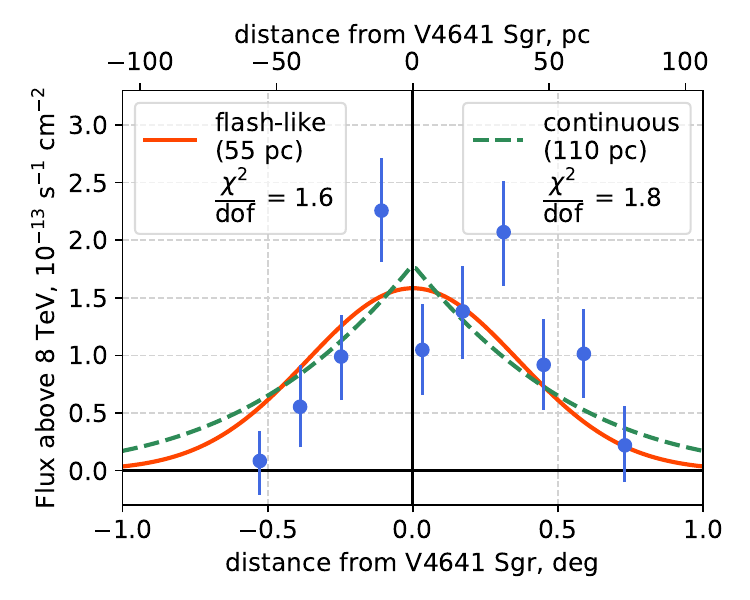}
    \caption{H.E.S.S. total flux above $0.8\dm{TeV}$ measurement along the major axis of the source~\cite{HESS-AA2025} fitted by flashlike (solid red) and continuous profiles (dashed green).} 
    \label{fig:profiles}
\end{figure}

Relying on the H.E.S.S. flux profile measurement along the major axis~\cite{HESS-AA2025}, we fit it with two models in \autoref{fig:profiles}: assuming a $t\dm{kyr}$ old single-flash scenario, and continuous injection from the source that started $t\dm{kyr}$ ago (see Appendix~\ref{appendix:1dprofiles}). The particles diffuse along the GMF with longitudinal diffusion coefficient $D$. Our fitting prefers \mbox{$\sqrt{4Dt}|^\rm{flash}_\rm{H.E.S.S.} = 55\dm{pc}$} for a single-flash case and \mbox{$\sqrt{4Dt}|^\rm{cont}_\rm{H.E.S.S.} = 110\dm{pc}$} for a continuous injection case. Under the assumption of $\eta = 1$ and \mbox{$B=4\dm{\mu G}$}, these values correspond to ages of \mbox{$t_\rm{flash}=2\dm{kyr}$} and \mbox{$t_\rm{cont}=8\dm{kyr}$} respectively~\cite{Semikoz-MF}.

The timescales derived above are lower bounds, since our calculation assumes that the nebula is oriented perpendicular to the line of sight (LOS). If, however, the major axis of the nebula is inclined at an angle $\theta < 90^\circ$ relative to the LOS, the time scales increase by a factor of $1/\sin^2\theta$.

We notice that the continuous emission profile (dashed green line in \autoref{fig:profiles}) is more peaked in comparison to the flashlike case (solid red line) . Hence, a more precise profile measurement might allow for distinguishing between these two scenarios and put constraints on the acceleration region.

In quasilinear theory with Kolmogorov turbulence spectrum and $\eta\ll 1$, the steady-state diffusion coefficient \mbox{$D_\parallel\propto (E/B)^{1/3} \eta^{-2}$}, which means that expected time increases with magnetic field and turbulent/regular field ratio~\cite{Reichherzer2020}. For $\eta\sim1$, diffusion coefficient behavior remains qualitatively similar~\cite{Semikoz-MF}. Anisotropic turbulent field simulations show that transverse diffusion coefficient $D_\perp$ in the range of our interest scales in the same way with energy and magnetic field, thus conserving the ratio $D_\parallel/D_\perp$~\cite{Semikoz-MF, Mertsch2020}.

Using the orthogonal extension of the nebula \mbox{$\sigma_\rm{minor}=0.06^\circ\pm0.01^\circ_\rm{stat}\pm0.01^\circ_\rm{sys}$} reported by H.E.S.S.~\cite{HESS-AA2025}, one can deduce the turbulent/regular magnetic field ratio~$\eta$, following the approach of Ref.~\cite{Semikoz-MF}. For a large range of possible projection angles~$\theta$ we find $\eta \sim 1$.

\section{Leptonic Scenarios}
\label{section:leptonic}

\begin{figure*}
    \includegraphics[width=.28\linewidth]{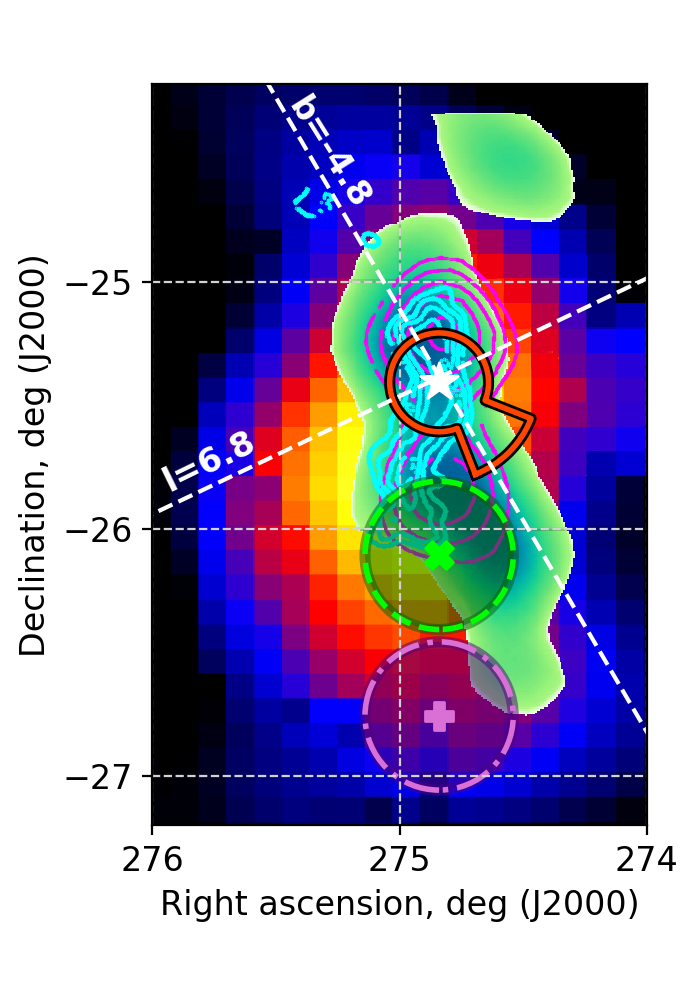}
    \includegraphics[width=.71\linewidth]{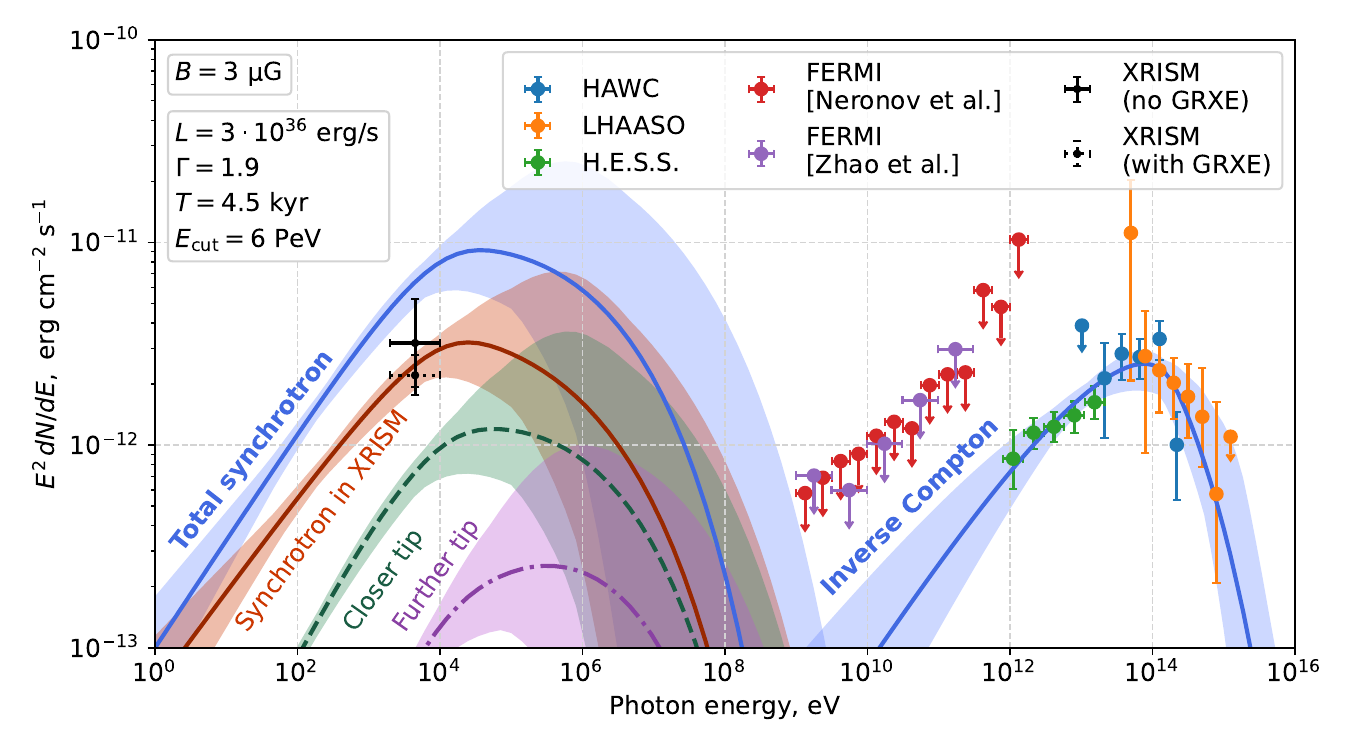}
    \caption{
    A \emph{continuous leptonic emission} model fitting the deabsorbed gamma-ray and x-ray emission measurements.\\
    \emph{Left panel:} the contours and TS maps of the source as observed by H.E.S.S.~\cite{HESS-AA2025}, HAWC~\cite{HAWC2024}, and LHAASO~\cite{LHAASO2024} in equatorial frame with the XRISM detection region~\cite{Suzuki2025} (red) and two proposed measurement regions (green and purple).\\
    \emph{Right panel:} photon spectrum, produced by continuous leptonic emission at the source for $B=3\dm{\mu G}$, $l_\parallel = 110\dm{pc},$ and $D_\parallel/D_\perp = 13$. Total spectrum (IC and Synchrotron emission) is shown in blue, the red synchrotron curve represents the flux from the electrons inside the XRISM region, the green and the purple curves correspond to the expected synchrotron flux at the edges of the nebula. The colors of the dashed curves correspond to the color of the regions in the left panel.
    }
    \label{fig:continuous_electrons}
\end{figure*}

In leptonic scenarios, we assume a power-law injection spectrum with a cutoff:
\mbox{$dN/dE = \rm{const}\times(E/E_0)^{-\Gamma}\,e^{-E/E_\mathrm{cut}}$}. The main photon production mechanisms are synchrotron radiation on~GMF and inverse Compton (IC) scattering on~CMB and~infrared photons. Although the model for background radiation we used is axisymmetric \cite{Popescu2017}, we find it applicable as~WISE and~Planck measurements do not reveal any particular radiation density features in~the~region of~the~source~\cite{Zhao2025}.

Despite its lower energy density, (\mbox{$\rho_\rm{CMB}\approx 0.26\dm{eV\,cm^{-3}}$}, \mbox{$\rho_\rm{dust}\approx0.72\dm{eV\,cm^{-3}}$}, \mbox{$\rho_\rm{starlight}\approx1.67\dm{eV\,cm^{-3}}$}) CMB remains dominant in the IC scattering. At high energies, IC scattering is in the Klein-Nishina regime, which makes background photon number density govern the interactions:  \mbox{$n_\rm{CMB}\approx410\dm{cm^{-3}}$}, \mbox{$n_\rm{dust}\approx60\dm{cm^{-3}}$}, \mbox{$n_\rm{starlight} \approx 2.7\dm{cm^{-3}}$} (see \autoref{fig:bg-rad} of Appendix~\ref{appendix:binary_lightcurve}).

In this section, our calculations are based on analytical expressions for a locally isotropic high-energy electron distribution \cite{Blumenthal_etal_1970}, as we assume that in the energy range of interest, diffusion has set in. We also use the parametrization of the total synchtrotron power spectrum function from~Ref.~\cite{Fouka_2013}.

\subsection{Single-flash leptonic scenario}

We begin with a~model of~an~instantaneous emission of~leptons, which happened during a major flash $t\dm{kyr}$ ago.

In this \emph{single-flash leptonic scenario}, the main constraint on the high-energy nebula's lifetime comes from synchrotron cooling of electrons in the GMF. The maximum observed photon energy of $\approx 800\dm{TeV}$~\cite{LHAASO2024} implies $t < 16\dm{kyr}\,(B/\rm{\mu G})^{-2}$. The minimal possible field strength needed to satisfy XRISM x-ray measurements~\cite{Suzuki2025} via synchrotron emission of relativistic electrons is $2\dm{\mu G}$, assuming all radiation originates within the XRISM region. Rescaling to the measurement region raises the minimum magnetic field to $5\dm{\mu G}$, reducing the lifetime estimate to \mbox{$t < 640~\mathrm{yr}$}. To match the high-energy LHAASO data, in a Markov Chain Monte Carlo (MCMC) parameter sampling with \texttt{emcee} Python library~\cite{emcee2013}, we encountered an even stricter constraint, $t < 450~\mathrm{yr}$. Both these times are insufficient for particles to reach the nebula’s edge even in the free streaming regime. \emph{Thus, the single-flash leptonic scenario is disfavored.}

\subsection{Continuous leptonic emission}

To relax the electron cooling constraint, we assume continuous emission of high-energy electrons from the source. This could correspond either to a persistent jet or to frequent flares, such as the one that occurred in~1999.

MCMC sampling shows that the minimal magnetic field to account for both x-ray and high-energy components is \mbox{$B=3\dm{\mu G}.$} To treat XRISM measurements correctly, we rescaled the synchrotron spectrum using our best-fit continuous emission diffusive profile from section~\ref{section:propagation}: \mbox{$\sqrt{4D_\parallel t}|_\rm{H.E.S.S.} = 110\dm{pc}$} and $D_\parallel/D_\perp = 13$. The corresponding spectra in gamma- and x-rays are presented in the right panel of \autoref{fig:continuous_electrons}. The synchrotron flux in the continuous leptonic emission scenario is also consistent with the recent MeerKAT measurements of the flux in radio range as presented in Ref.~\cite{Grollimund2026}.

From an energetic perspective, continuous emission leptonic scenarios are reasonable: the luminosity required to power the large-scale electron nebula varies between $10^{36}$ and $10^{38}\dm{erg\,s^{-1}}$, which is less than the black hole Eddington luminosity  $L_\mathrm{Ed} \approx 10^{39}\dm{erg\,s^{-1}}$. In magnetic field of $3\dm{\mu G}$, the luminosity required to power the electron population reaches a minimum of $2\times10^{36}\dm{erg\,s^{-1}}$ for a nebula lifetime of $\sim 7\dm{kyr}$ (see Appendix~\ref{appendix:leptonic-luminosity}).

From a morphological standpoint, however, leptonic scenarios are challenging. Since the background radiation consists of the CMB, dust thermal radiation, and starlight, it is not expected to vary significantly on the scales of the nebula. Given that the main scattering targets are CMB photons, the asymmetry can be explained only by electron transport effects.

\subsection{Asymmetry in leptonic scenarios}

As mentionned in Sec.~\ref{section:mq}, the high-energy $\gamma$-ray TS maps show the nebula's asymmetry, which grows with energy. If H.E.S.S. observations are almost symmetric, the high-energy HAWC and LHAASO maps demonstrate that the southern edge of the nebula is further from the source than the northern one.

We propose two possible mechanisms to explain the observed asymmetry in the leptonic scenario: (a) bending of the GMF lines near the source, and (b) ballistic-diffusive transition of high-energy particles.

\begin{figure}[t]
    \centering
    \includegraphics[width=\linewidth]{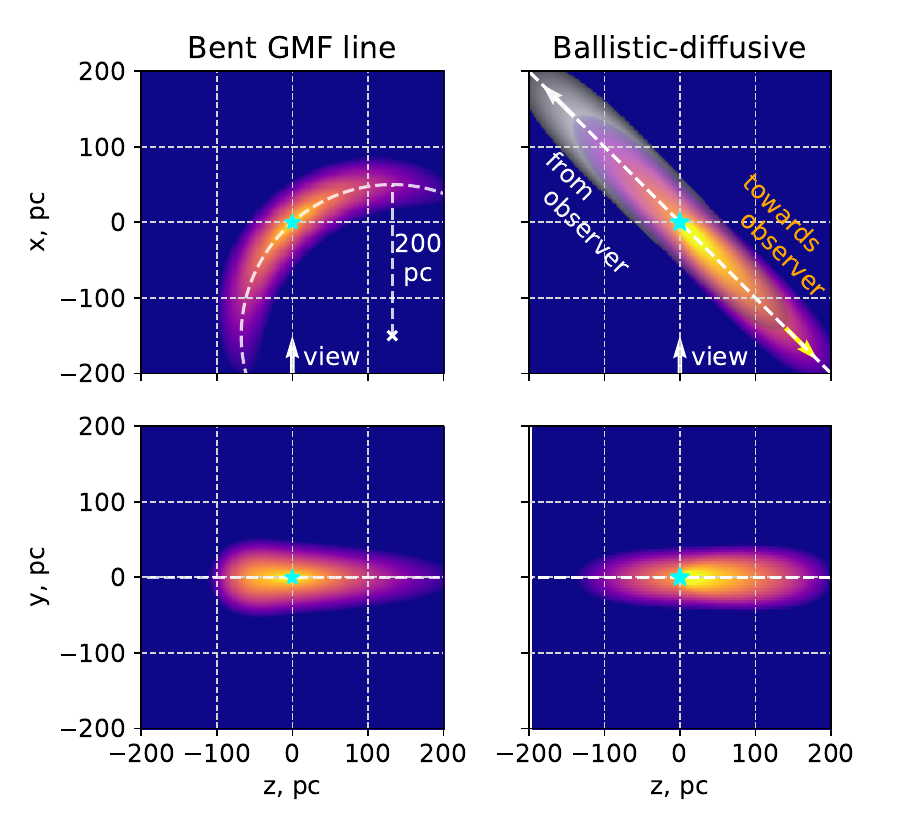}
    \caption{
    Two leptonic scenarios which can account for the observed asymmetry of the nebula. The $x$-axis coincides with the line-of-sight; $z$-axis is orthogonal and corresponds to the projection of the nebula major axis onto the celestial sphere. In all panels, the source is marked as a blue star at the origin. The color map corresponds to the lepton number density, which is proportional to the expected IC gamma-ray flux.\\
    \emph{Left column:} top and side views of the nebula in the ``bent GMF line'' case, the curvature radius is $200\dm{pc}$, the curvature center is marked with a white cross.\\
    \emph{Right column:} top and side view of the nebula in the ``ballistic-diffusive transition'' scenario. In the top plate, the projection angle is $45^\circ$, the particles coming towards the observer are presented in yellow-red color map, the ones propagating from the observer are marked in gray. In the lower plate, only the particles coming towards the observer are shown.  
    }
    \label{fig:two_scenarios}
\end{figure}

\begin{enumerate}[nosep=false]
    \item \emph{If the GMF lines are bent, }in order to acquire the observed asymmetry, the curvature radius of the field line has to be of the order of $R_\rm{curv}\sim 200\dm{pc}$. Since diffusion coefficients depend on energy, low-energy electrons would spread out at shorter distances, thus resulting in smaller asymmetry level, at higher energies, when $\sqrt{4Dt}$ approaches $ R_\rm{curv}$, asymmetry is the strongest. Although this scenario provides an explanation for the observed north-south asymmetry of the source, it also requires the northern part to be brighter, which is not clearly supported by the existing data. The top view and the observed nebula profiles in the bent-GMF-line scenario are presented in the left column of~\autoref{fig:two_scenarios}.

    \item \emph{In the ballistic-diffusive transition scenario}, we consider the spatial diffusion to be not completely established and assume pitch-angle diffusion as a transport regime between free streaming along a GMF line and spatial diffusion in the anisotropic turbulent GMF. Since the diffusion coefficient depends on energy, low-energy particles enter the diffusive regime earlier than high-energy ones. Therefore, if the low-energy emission is already symmetric, high-energy particles may still exhibit a preferred propagation direction, making the near edge of the nebula appear brighter than the far edge.\\    
    Inspired by Ref.~\cite{Tautz2016}, we model the anisotropic ballistic-diffusion transition using the one-dimensional \emph{telegrapher’s equation} along the GMF line, \mbox{$\left(\partial^2_{tt} + \tau^{-1} \partial_t - v_0^2\partial^2_{zz}\right)n(z,t) = \left(\partial_t + \tau^{-1}\right)q(z,t),$} which describes free streaming of particles with velocity $v_0$ for $t \ll \tau$ and spatial diffusion with $D_\parallel = \tau v_0^2$ for $t \gg \tau$. Here, $n(E,t)$ is the particle density, and $\tau$ is the scattering time. To construct a three-dimensional solution, we combine the one-dimensional Green’s function of the telegrapher’s equation with an orthogonal diffusion kernel (see Appendix~\ref{appendix:telegraphers-greens-function}). The top view and observed nebula profiles in ballistic-diffusive transition scenario, are shown in the right column of the \autoref{fig:two_scenarios}. 
\end{enumerate}

\section{Leptohadronic Scenarios}
\label{section:leptohadronic}

Since the gamma-ray background is CMB dominated, the energy threshold to $p\gamma$ interaction is much higher than $1\dm{PeV}$. Hence, we consider that in hadronic scenarios gamma-ray production is governed by proton-proton interactions. In this process, the average photon energy is $\sim 20$ times less than that of the parent proton. The efficiency of pion production depends on the background proton density. Axisymmetric~\cite{Misiriotis2006} and nonsymmetric~\cite{Ferrire2007} Milky Way central region gas models predict the total proton density at the source position to be extremely low: $n_\rm{S} \approx 0.006\dm{cm^{-3}}$ and $n_\rm{NS} \approx 0.03\dm{cm^{-3}}$ respectively. Density enhancements may come from overdense regions that are not accounted for in the global gas distribution models. To search for possible local density enhancements, we studied the HI density directly from the neutral hydrogen all-sky survey HI4PI~\cite{HI4PI2016}.

\subsection{Cold hydrogen map}

\begin{figure}[t]
    \centering
    \includegraphics[width=1.04\linewidth]{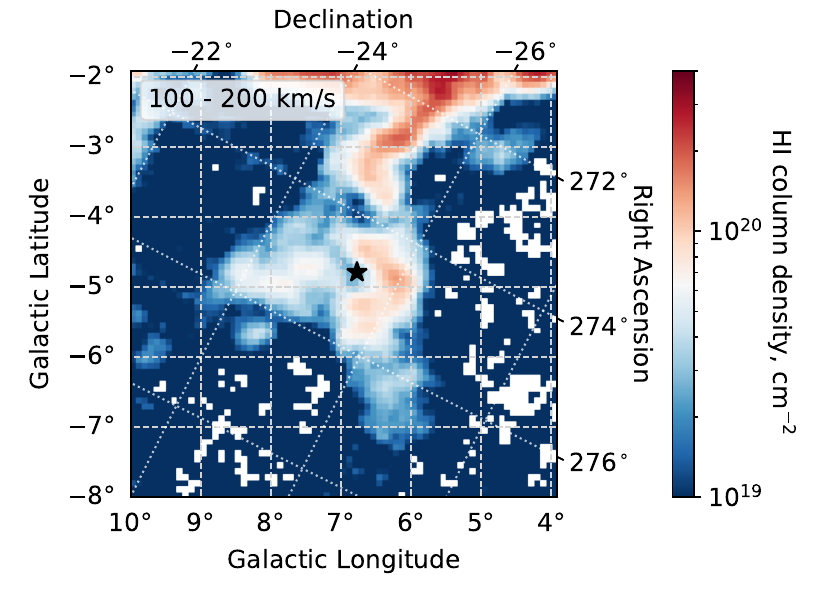}
    \vspace{-2ex}
    \caption{
    HI column density in the vicinity of the source in the velocity domain $100 < v < 200\dm{km/s}$~\cite{HI4PI2016}. A moderate overdensity region is seen around the source.}
    \vspace{-2ex}
    \label{fig:HI-averaged-map}
\end{figure}

In the vicinity of the Galactic bar, the gas is strongly perturbed by its gravitational potential, and the standard tangent-point method for connecting distances and velocities is inapplicable \cite{WrongRotationCurve-2015}. Hydrodynamical simulations of the gas in the Milky Way suggest that the corresponding radial gas velocities in the region of V4641~Sgr are \mbox{$100< v < 200 \dm{km/s}$} (see Fig.\,4 from \cite{Li2022}). \autoref{fig:HI-averaged-map} shows a local HI overdensity, which extends approximately $1.5^\circ$ (160~pc at $d=6.1\dm{kpc}$) around the source. By assuming this region to be spherically symmetric, we estimate the average HI density to be $n_\rm{HI}\sim 0.1\dm{cm^{-3}}$, which increases to $1\dm{cm^{-3}}$ in the most dense regions.

From the map in \autoref{fig:HI-averaged-map}, we observe an asymmetry in the density that coincides with the direction of the source: in the lower right (southern) region, the density is higher than in the upper left (northern) one. Therefore, if the detected overdensity corresponds to the source location, the asymmetry observed by HAWC and LHAASO can be directly explained by the enhanced gas density on one side of the source.

\begin{figure*}
    \centering
    \vspace{-2ex}
    \includegraphics[width=\linewidth]{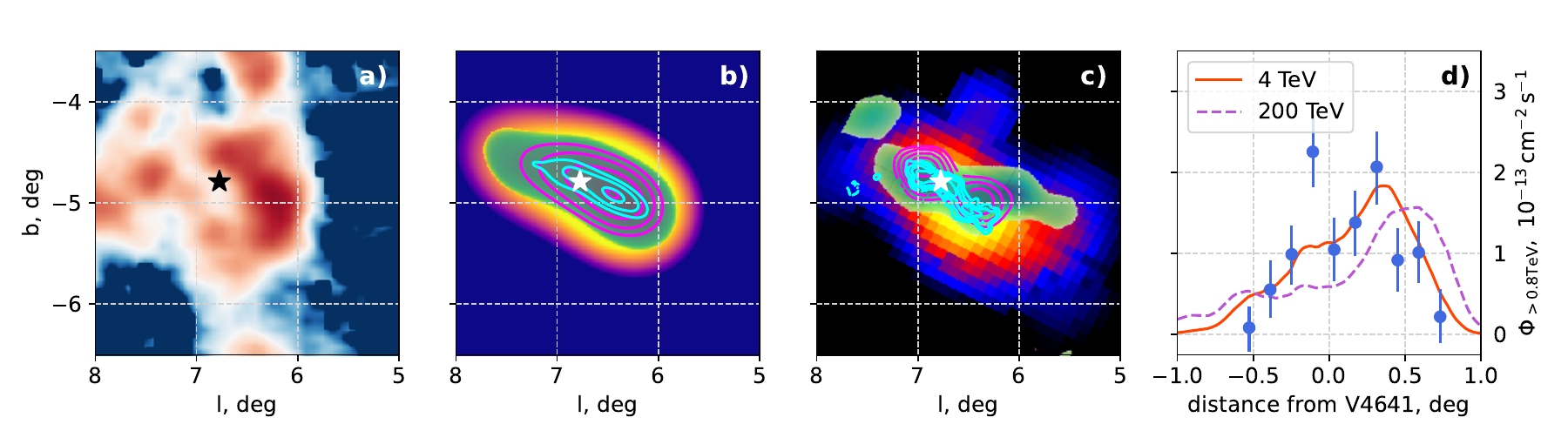}
    \vspace{-4ex}
    \caption{
    \emph{Panel a):} HI column density map (log scale) from HI4PI survey at $v=125\dm{km/s}$~\cite{HI4PI2016}.\\\emph{Panel b):} predicted hadronic emission pattern in single hadronic flash diffusive scenario: yellow-red pattern corresponds to LHAASO measurements~\cite{LHAASO2024}, green-blue one stands for HAWC $>100\dm{TeV}$, magenta contours correspond to overall HAWC result~\cite{HAWC2024}, and cyan contours are based on H.E.S.S. data.~\cite{HESS-2024}.\\
    \emph{Panel c):} measured gamma-ray TS contours and TS maps (notation same as in \autoref{fig:combined_image}).\\
    \emph{Panel d):} flux profile H.E.S.S. measurements (blue points)~\cite{HESS-AA2025}, diffusion profile at H.E.S.S. mean energy of 4~TeV (solid red) and at LHAASO mean energy of~200~TeV (dashed purple).
    }
    \label{fig:map_comparison}

    \includegraphics[width=\linewidth]{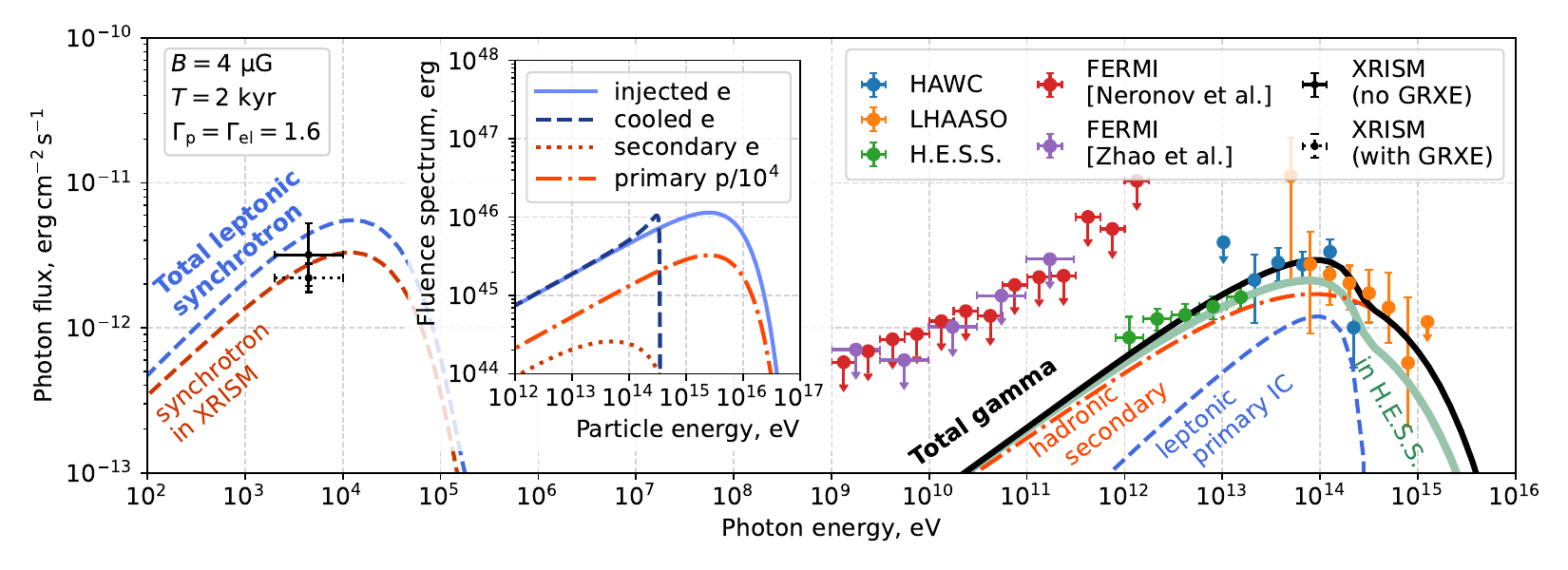}
    \vspace{-6ex}
    \caption{
    \emph{Main figure:} deabsorbed extended gamma-ray measurements~\cite{HESS-AA2025, HAWC2024, LHAASO2024, Neronov2024, Zhao2025} and x-ray points~\cite{Suzuki2025} with \emph{leptohadronic flash} model curves: red dash-dotted curve corresponds to hadronic secondary gamma-rays; blue dashed one stands for leptonic IC and total synchrotron radiation; solid black curve shows total gamma-ray radiation; light green solid curve shows the gamma-ray flux in the H.E.S.S. measurement region.\\
    \emph{Inset figure:} spectra of injected and cooled electrons in solid light blue and dashed white respectively; scaled by factor of~$10^4$ primary protons, and secondary electrons in dotted light red.
    }
    \label{fig:leptohadr_flash}

    \includegraphics[width=\linewidth]{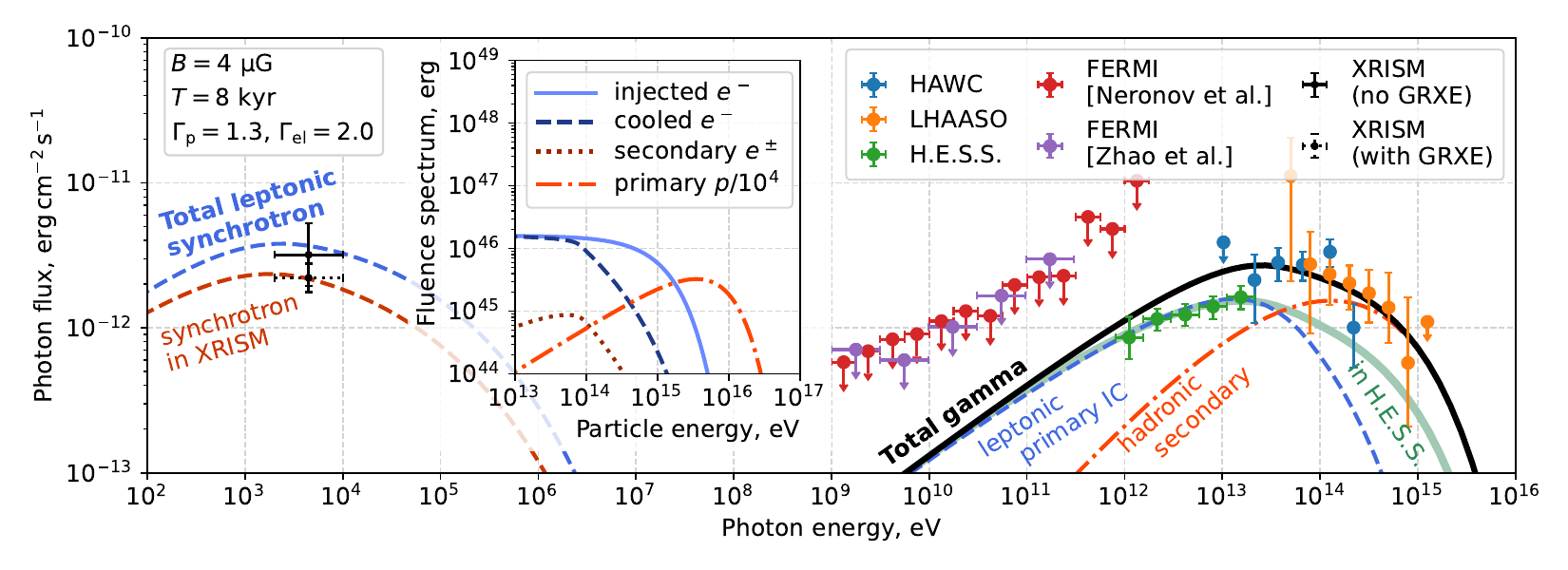}
    \vspace{-4ex}
    \caption{Experimental data and \emph{hadronic flash accompanied by continuous leptonic emission model} curves in notation of \autoref{fig:leptohadr_flash}.
    }
    \label{fig:hadronic_flash_continuous_leptonic}
\end{figure*}

\autoref{fig:map_comparison} shows the HI column density map and the predicted hadronic flux of the source compared to the H.E.S.S., HAWC and LHAASO signal maps. The map of~\emph{panel~b)} in~\autoref{fig:map_comparison} is constructed as follows:

\begin{enumerate}[label={(\arabic*)}]
    \item We took $l_\parallel = 55\dm{pc}$ and $D_\parallel/D_\perp=13$ from the H.E.S.S. measurement fit at $4\dm{TeV}$;

    \item We assumed a Kolmogorov turbulence spectrum ($D_\parallel,D_\perp\propto E^{1/3}$, which implies $l = \sqrt{4Dt} \propto E^{1/6};$

    \item We calculated diffusion scales for HAWC, HAWC~$>100\dm{TeV}$, and LHAASO at corresponding energies of $10\dm{TeV},~100\dm{TeV},~200\dm{TeV}$ and computed diffusive proton distribution profiles along a straight magnetic field line, which has a $25^\circ$ inclination to the line of constant Galactic latitude; 

    \item As $\gamma$-ray intensity is proportional to both low- and high-energy proton densities \mbox{$I_\gamma \propto n_\rm{cold}\times n_\rm{HE}$}, we computed the expected flux maps and approximate contours at the characteristic energies of H.E.S.S., HAWC, and LHAASO, assuming high-energy proton emission symmetric about the source.
\end{enumerate}

\subsection{Hadronic flash}
\label{subsection:hadronic}

A purely hadronic scenario was discussed in Ref.~\cite{Neronov2024}. The parameters of the proposed proton spectrum were $\Gamma=1.8$, $E_\rm{cut} = 5\dm{PeV},$ total energy stored in protons for this scenario is $\mathcal{E}_\rm{tot} = 2\cdot10^{50}(n_0/n)\dm{erg}$, where $n_0 = 1\dm{cm^{-3}}$ and $n$ is the average cold proton density in the region.

XRISM x-ray measurements strongly overshoot the predicted synchrotron flux from secondary electrons, which allows two interpretations. It is in principle possible that the x-ray flux does not correlate with the high-energy emission form the source. Indeed, the inner Galaxy region is known to host x-ray bubbles of complex morphology \cite{Predehl:2020kyq} and the emission observed by XRISM may be part of a larger x-ray structure unrelated to the source. Otherwise, if the x-ray emission is associated to the source, the synchrotron emission from secondary electrons produced in proton-proton collisions is insufficient to explain the x-ray data and there exists a population of directly accelerated electrons that produces the observed x-ray flux.

To account for the XRISM measurement, we constructed two leptohadronic models, which explain both the morphological and spectral properties of the source. To treat secondary particles in $pp$~interactions, we used the \texttt{aafrag} Python library~\cite{aafrag2021}.

\subsection{Leptohadronic flash}
\label{subsection:leptohadronic_flash}

The first model assumes a single flash of protons and electrons, accelerated  $t\dm{kyr}$ ago.

The electrons are cooling through synchrotron and inverse Compton energy loss. This places constraints on electron population's lifetime: the synchrotron radiation peak has to remain within or above the XRISM energy range \mbox{(2-10$\dm{keV}$).} The synchrotron radiation spectrum peaks at \mbox{$\epsilon_\rm{synch} \simeq 
20\dm{keV}\,\left(E/\rm{PeV}\right)^2\left(B/\rm{\mu G}\right)$}; maximal electron energy after time $t$ is \mbox{$E_\rm{max} = 13\dm{PeV}\,\left(B/\rm{\mu G}\right)^{-2}\left(t/\rm{kyr}\right)^{-1}$}. This leads to a constraint on the source age \mbox{$t < 31\,\rm{kyr}\,\left(B/\rm{\mu G}\right)^{-3/2}.$}

An exploration of MF strength range from $1$ to $10\dm{\mu G}$ shows that minimal magnetic field is $4\dm{\mu G}$ and thus gives even shorter lifetime $t < 4\dm{kyr}$, consistent with the best-fit flashlike transport age \mbox{($t_\rm{flash} = 2\dm{kyr}$)} obtained in Sec.~\ref{section:propagation}.

In this scenario, leptonic and hadronic particles are assumed to be accelerated in the same region during a single major event. We use a simplified injection model with identical spectral indices and a common high-energy cutoff, reducing the number of independent parameters. The multiwavelength data are well described within this leptohadronic framework. The corresponding parameters are summarized in Table~\ref{tab:leptohadronic_parameters}.

In \autoref{fig:leptohadr_flash}, we present the total predicted gamma-ray flux, and the hadronic and leptonic contributions separately, and the portion of the total flux seen in the H.E.S.S. measurement region ($r < 55\dm{pc}$). Total synchrotron emission is shown along with its fraction in the XRISM region. A~contribution from secondary electrons in the model is negligible, as shown in the inset plot of the figure.

\begin{table}[h!]
    \centering
    \caption{Possible parameters of the leptonic and hadronic populations in the \emph{leptohadronic flash model.} $n_0 = 1\dm{cm^{-3}}$, and $n$~is~the average cold proton density in the region.}
    \vspace{2ex}
    \label{tab:leptohadronic_parameters}
    \begin{tabular}{l|c|c|c|c|c}
        \hline
        & $B,\rm{\mu G}$ & $T,\rm{kyr}$ & $\Gamma$ & $E_\rm{cut},\rm{PeV}$ & $\mathcal{E}_\rm{tot},\dm{erg}$ \\
        \hline
        Hadronic   & \multirow{2}{*}{4}  & \multirow{2}{*}{2.6} & \multirow{2}{*}{\,1.6\,} & \multirow{2}{*}{5} & $1.5\times10^{50}(n_0/n)$ \\
        Leptonic   & &&&& $0.5\times10^{47}$ \\
        \hline
    \end{tabular}
\end{table}

\subsection{Hadronic flash and leptonic emission}
\label{subsection:hadronic_flash_leptonic_emission}

As done previously for the leptonic model, to weaken the synchrotron cooling constraint, we assume a hadronic flash $t\dm{kyr}$ ago, accompanied by continuous leptonic emission.

We conclude that a $4\dm{\mu G}$ magnetic field strength allows for a self-consistent spectral description of the nebula. A~possible model for this scenario has a lifetime of $t = 8\dm{kyr}$, consistent with the best-fit timescale for the continuous emission (Sec.~\ref{section:propagation}). The model parameters are summarized in~\autoref{tab:hadronic_flash_leptonic_emission_parameters}, and the corresponding spectral fit is presented in \autoref{fig:hadronic_flash_continuous_leptonic}. The required electron injection luminosity is \mbox{$L_\rm{el} = 0.8\times10^{36}\dm{erg/s} < 10^{-3}\times L_\rm{Ed}.$}

\begin{table}[h]
    \centering
    \caption{Possible parameters of the leptonic and hadronic populations in the \emph{hadronic flash and leptonic emission model.} $n_0 = 1\dm{cm^{-3}}$, and $n$~is~the average cold proton density.}
    \vspace{2ex}
    \label{tab:hadronic_flash_leptonic_emission_parameters}
    \begin{tabular}{l|c|c|c|c|c}
        \hline
        & $B,\rm{\mu G}$ & $T,\rm{kyr}$ & $\Gamma$ & $E_\rm{cut},\rm{PeV}$ & $\mathcal{E}_\rm{tot},\dm{erg}$ \\
        \hline
        Hadronic   & \multirow{2}{*}{4}  & \multirow{2}{*}{8} & \,1.3\, & 5 & $0.8\times10^{50}(n_0/n)$ \\
        Leptonic   & & & \,2.0\, & 1 & $2.1\times10^{47}$ \\
        \hline
    \end{tabular}
\end{table}

These parameters correspond to the \emph{minimal magnetic field} compatible with the data, and \emph{the fit is not unique}. For instance, adopting a stronger magnetic field ($10\dm{\mu G}$) allows for a softer proton spectrum ($\Gamma_\rm{prot}\approx 1.6$) but requires hardening of the electron spectrum ($\Gamma_\rm{el}\approx 1.8$) to compensate for enhanced synchrotron cooling. In this case, the leptonic contribution to gamma rays shifts to lower energies, and signal registered in H.E.S.S. also becomes hadronic-dominated. Total energy in protons increases to $1.7\times10^{50}(n_0/n)\dm{erg},$ and electron luminosity decreases to $0.5\times10^{36}\dm{erg/s}$.

As shown in~\cite{Grollimund2026}, the MeerKAT radio flux measurements and the XRISM x-ray measurement require the electron power-law index to be $\Gamma_\rm{el} \sim 2.1$ to allow a synchrotron-radiation-based explanation, which is consistent with our results. If both radio and x-ray extended emissions correspond to the same high-energy electron population, this might impose additional constraints on the magnetic field in leptohadronic scenarios. Extension of the synchrotron spectrum presented in~\autoref{fig:hadronic_flash_continuous_leptonic} to the radio domain, gives a mild underprediction of the MeerKAT  flux by a factor of~$\sim2$.

In both scenarios (\ref{subsection:leptohadronic_flash},~\ref{subsection:hadronic_flash_leptonic_emission}), the gamma-ray data require \emph{intrinsically hard hadronic spectra} with $\Gamma = 1.6$ and $\Gamma = 1.3$ respectively. Such hard spectra may arise under extreme conditions during the injection episode, for example via magnetic reconnection acceleration~\cite{Sironi2014, Guo2014}, turbulence acceleration in highly magnetized turbulence~\cite{Rieger2004, Comisso2019}, or particle acceleration in vacuum gaps~\cite{Neronov2003, Ptitsyna2016}. In~(\ref{subsection:hadronic_flash_leptonic_emission}), electrons may be injected continuously, during episodic flares, or during active phases of the binary. Given the observed source activity (Appendix~\ref{appendix:binary_lightcurve}), both possibilities are energetically viable.

We emphasize that in the presented leptohadronic flash model~(\ref{subsection:leptohadronic_flash}) gamma-ray emission is dominated by hadronic processes in H.E.S.S., HAWC and LHAASO. On the contrary, in the hadronic flash and leptonic emission model~(\ref{subsection:hadronic_flash_leptonic_emission}) the lower-energy part of the gamma-ray spectrum ($E < 40\dm{TeV}$) consists mostly of the IC photons, the higher-energy part ($E > 100\dm{TeV}$) contains secondary photons from $\pi^0$ decay from $pp$ interactions. Thus, due to energy dependence of diffusion coefficient, average size of hadronic nebula is several times larger than the corresponding scales of the leptonic nebula.

\subsection{Neutrino flux}

\begin{figure}[h]
    \centering
    \includegraphics[width=\linewidth]{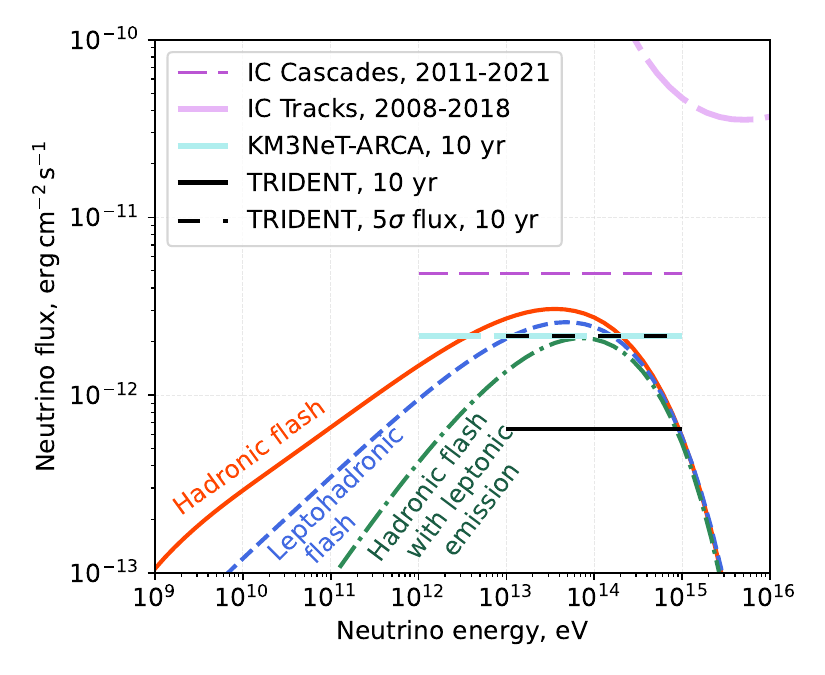}
    \caption{Predicted all-flavor neutrino fluxes from the hadronic (red solid), leptohadronic flash (blue dashed), and hadronic flash with leptonic emission (green dash-dotted) models compared to the sensitivities of the existing detectors IceCube (IC) and KM3NeT-ARCA, the proposed TRIDENT detector $90\%$ and $5\sigma$ sensitivities are shown in dashed and solid black respectively~\cite{Neronov2024}.}
    \label{fig:neutrinos}
\end{figure}

Hadronic interactions, if present at the source, leave a detectable neutrino footprint. We calculated expected neutrino fluxes for the source with the use of \texttt{aafrag}~\cite{aafrag2021}. In \autoref{fig:neutrinos}, we show the predicted neutrino fluxes from the hadronic model (\ref{subsection:hadronic}) from~Ref.~\cite{Neronov2024} and a leptohadronic scenarios (\ref{subsection:leptohadronic_flash}) or (\ref{subsection:hadronic_flash_leptonic_emission}). We also compare them with the sensitivities of the existing facilities: IceCube~\cite{IceCube:2023ame}, KM3NeT-ARCA~\cite{Aiello2019, KM3Net} and the proposed TRIDENT~\cite{TRIDENT} detector. Expected sensitivity of the HUNT project~\cite{HUNT} should be at the same level as the TRIDENT experiment assuming equivalent detector volume.

We accounted for the angular resolution of IceCube in track channel, $\theta \simeq 0.2^\circ$ is better than the source size and the source would be seen as $\simeq 5$ pointlike sources so that the flux of each source is $\simeq 1/5$ of the total. We assume for all neutrino telescope a limiting angular resolution of $0.2^\circ$ for tracks and apply the same penalty factor as for IceCube tracks. \vspace{-1ex}

\vspace{-3ex}

\section{Discussion and Conclusion}
\label{section:discussion}

In the present work, we discuss leptonic, hadronic and leptohadronic models of the high-energy nebula, which was observed successively by HAWC, LHAASO, and H.E.S.S. The combination of the signal maps corresponding to these observations is shown in \autoref{fig:combined_image}. The deabsorbed spectral properties of the source measured by the aforementioned experiments with the corresponding Fermi-LAT upper limits are presented in \autoref{fig:continuous_electrons}, \autoref{fig:leptohadr_flash}, and \autoref{fig:hadronic_flash_continuous_leptonic}. XRISM analysis~\cite{Suzuki2025} shows the measurement of extended x-ray emission around the binary, the corresponding measurement region is also shown in \autoref{fig:combined_image}. Recently, an extended cloud around the source was found in radio domain by MeerKAT~\cite{Grollimund2026}.

Various models for the source have already been considered in recent works on V4641~Sgr: a purely leptonic model based on shear acceleration in the jet can be found in~\cite{leptonic2025}, a turbulence acceleration model is developed in~\cite{Dmytriiev2025}; a purely hadronic model is presented in~\cite{Neronov2024}; a leptohadronic model, which accounts for both gamma-ray and x-ray extended radiation is constructed in \cite{leptohadronic2025}.

In the present work, we provide a self-consistent description of the high-energy nebula, which addresses both spectral and morphological properties of the source.

Energetically, purely leptonic model is preferred as it allows for explaining the measurements with a continuous emission of a moderately powerful jet $<0.01 L_\rm{Ed}.$ However, morphological properties of the gamma-ray region in the leptonic model require peculiar assumptions (\autoref{fig:two_scenarios}), such as strong magnetic field line curvature or fine-tuned timescale and regular/turbulent GMF ratio to observe the ballistic-diffusion transition.

Hadronic and leptohadronic scenarios, on the contrary, allow for a simple proton-background-based morphology explanation. As the source is located in the vicinity of the Galactic bar (\autoref{fig:source_position}), the gas around it is perturbed, which requires considering a velocity range, different from the one given by a standard tangent point method. Using the radial gas velocity value of $125\dm{km/s}$, we modeled a possible hadronic contribution to the nebula, which allowed us to recover the asymmetry of the observed nebula (\autoref{fig:map_comparison}).

Nevertheless, as average cold proton density does not exceed $1\dm{cm^{-3}}$, from energetic point of view, the energy stored in high-energy protons is enormous $E > 10^{50} \dm{erg}$, which could be released only in a catastrophic event, such as the infall of a large portion of the companion star onto the black hole, or the very formation of the black hole from a massive progenitor. XRISM x-ray measurement, if attributed to the high-energy particle population, makes the purely hadronic model inconsistent and requires an addition of a population of electrons for synchrotron radiation to occur.

To distinguish between the purely leptonic and leptohadronic scenarios, one can try measuring neutrino fluxes, as proposed in \cite{Neronov2024}. However, with the sensitivity of current detectors, it will take at least a decade to reach the required precision. A recent search of the signal from V4641~Sgr in 10 years of the IceCube data resulted into weak upper limits, which do not allow for making a conclusion on the origin of the gamma-ray emission~\cite{li2025IceCube}.

A more straightforward way to distinguish between the models is to conduct x-ray measurements at the tips of the nebula, which can be done by analyzing archival data of INTEGRAL~\cite{winkler2003integral}, or by making a new measurement by XRISM~\cite{Tashiro2020} and SVOM~\cite{Atteia2021}. In this work, we make a prediction for the minimal x-ray flux in the purely leptonic model (\autoref{fig:continuous_electrons}). 

An observation of the nebula's morphology at energies different from H.E.S.S. can be made by LACT, a future atmospheric Cherenkov telescope array at the LHAASO site with a comparable angular resolution of $0.1^\circ$~\cite{LACT2024, LACT2025V4641}, and by CTAO-South with better angular resolution and higher flux sensitivity~\cite{CTAO}. Detailed observations would allow for checking particle transport hypothesis, namely to study the length and width of the filament and their energy dependencies. 

    \section*{Acknowledgments}

This work  has received in part support under the program ''~Investissement d’Avenir~'' launched by the French Government and implemented by ANR, with the reference ''~ANR-18-IdEx-0001~'' as part of
its program ''~Emergence~''. We thank Laura Olivera Nieto and Kathryn Plant for valuable comments. 
    
    \bibliography{lit}
    
    \section{Appendix}
    \appendix
    \section{Binary lightcurve and spectrum}
\label{appendix:binary_lightcurve}

First mentioned in 1978, V4641 was incorrectly identified as a variable star GM Sgr, which is $1''$ away \cite{barsukova2014}. In February 1999, the binary was rediscovered as a faint X-ray variable source \cite{bepposax1999}. The most violent outburst of the source was observed in Radio, Optical, and X-ray domains in September 1999, when the X-ray flux reached 12 Crab units \cite{Hjellming2000, stubbings1999, 12Crab1999}. Despite the 2.8~day orbital period of the binary, the outburst variability scales were of the order of hundreds of seconds~\cite{Revnivtsev2002}.

After 1999, the source showed a number of less powerful outbursts in 2002, 2003~\cite{buxton2003atel170, bailyn2003atel171, rupen2003atel172}, 2004~\cite{swank2004atel295, rupen2004atel296, bikmaev2004atel309}, 2005~\cite{swank2005atel536, khamitov2005atel540}, 2010~\cite{yamaoka2010atel2785}, 2014~\cite{tachibana2014atel5803, uemura2014atel5836}, 2015~\cite{yoshi2015atel7858}, 2020~\cite{shaw2020atel13431, imazato2020atel13437, zhang2020atel13471}, 2021~\cite{negoro2021atel14968, goranskij2021atel15008}, and 2024~\cite{negoro2024atel16804, grollimund2024atel16852, goranskij2024atel16916}; after 2004 there were no radio outbursts reported until 2024~\cite{miller-jones2015atel7908, zhang2020atel13471, grollimund2024atel16852}, the source remained in a radio-quiet state for a long time. Faint X-ray outbursts happen rather regularly, slightly more rare than once a year. \autoref{fig:lightcurve} shows the source's historical lightcurve in visible (V) and blue (B) optical domains, based on~\cite{stubbings_blog, barsukova2014}; X-ray and radio flares are shown in pale violet and red lines.

In \autoref{fig:obs-spectrum}, we gathered several point-source flux measurements in a broad photon energy range, from radio to X-rays.  
\emph{Radio measurements} were taken in 1999~\cite{Hjellming2000} and 2004~\cite{rupen2004atel296}, with upper limits reported in 2003-2005~\cite{Pandey2007}, 2009~\cite{Miller-Jones2011}, 2015, 2020 and 2024 (\cite{miller-jones2015atel7908, zhang2020atel13471, grollimund2024atel16852}  respectively). 
\emph{Infrared and optical data} from the 1999 outburst are available in~\cite{Chaty2003}, blackbody radiation curve is drawn for \mbox{$T = 10\,500\dm{K}$}~\cite{Orosz2001}. 
\emph{X-ray observations} have been primarily conducted by NuSTAR~\cite{Connors2025} and Swift~\cite{Pahari2015}, with a key measurement of the 1999 flash provided by~\cite{Revnivtsev2002}. X-Ray flux in quiescence is registered with Chandra~\cite{Tomsick_2003}.
When relevant, the values were rescaled with the use of $d = 6.1\dm{kpc}.$

In \autoref{fig:bg-rad}, we show different contributions to background radiation field at the source location~\cite{Popescu2017}. Despite starlight dominates in energy density, to produce $>10\dm{TeV}$ photons, IC scattering has to be in KN regime, which reduces the interaction cross-section~\cite{crbeam}.

    \section{Size-on-energy dependence}
\label{appendix:size}

\begin{figure}[b!]
    \vspace{-3ex}
    \centering
    \includegraphics[width=\linewidth]{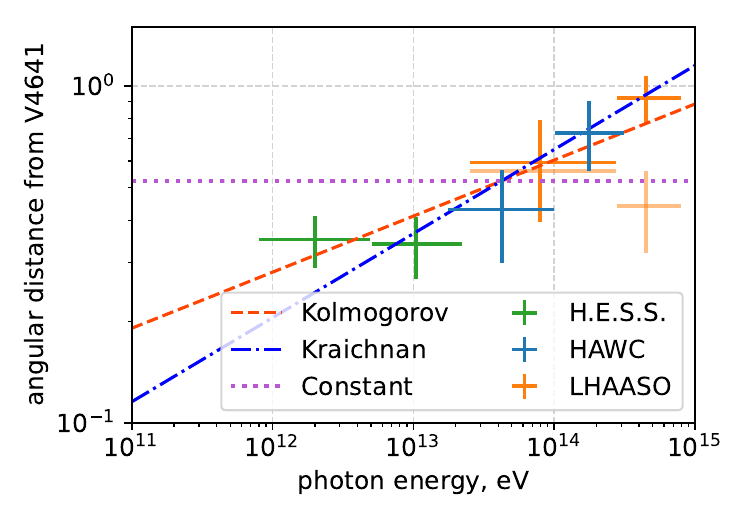}
    \vspace{-3ex}
    \caption{Size-on-energy dependence in H.E.S.S., HAWC, and LHAASO (inspired by \cite{HESS-AA2025}). Green and blue points show source extensions for H.E.S.S. and HAWC models respectively; semi-transparent orange points stand for extensions of LHAASO single-component elliptic models; opaque orange points show the distance of the Southern ellipsis edges from the binary~\cite{LHAASO2024}. Red, blue, and purple lines represent best-fit models of particle population scaling: Kolmogorov turbulence ($l\propto E^{1/6}$), Kraichnan turbulence ($l\propto E^{1/4}$), and energy-independent model respectively.}
    \label{fig:size-on-energy}
\end{figure}

\begin{figure*}[!b]
    \centering
    \includegraphics[width=\linewidth]{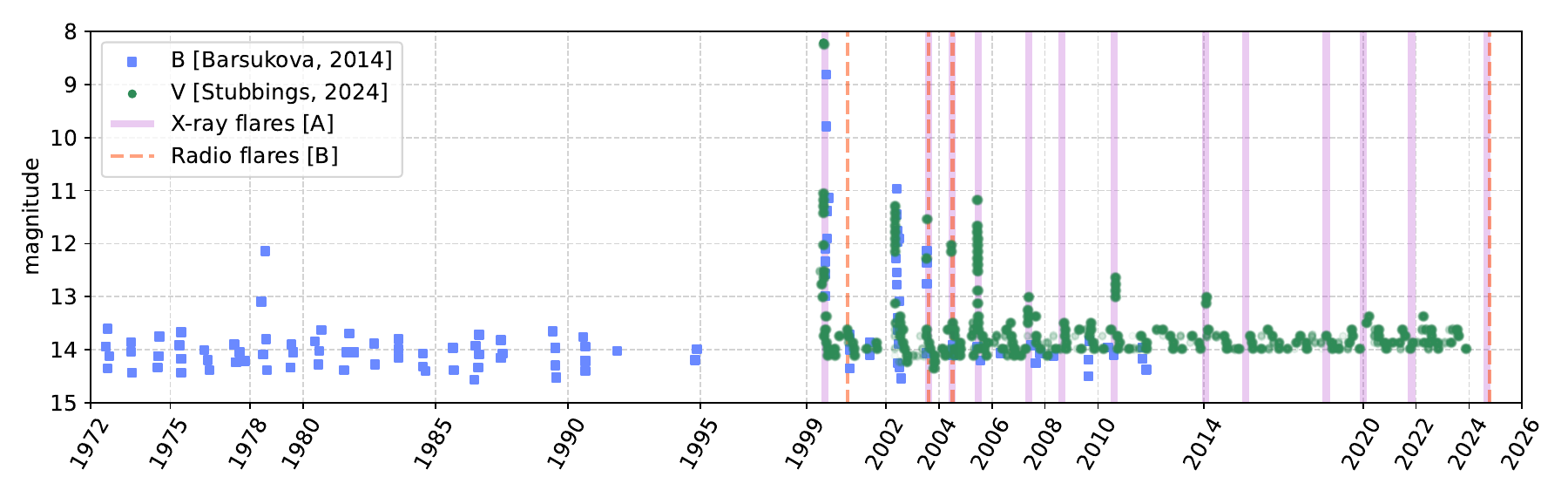}
    \caption{Historical lightcurve of the V4641 Sgr in blue~\cite{barsukova2014}, and visible~\cite{stubbings_blog} bands. X-ray flares are shown in pale violet lines, radio flares are marked with dashed red lines. The sources are specified in the text.}
    \label{fig:lightcurve}

    \includegraphics[width=\linewidth]{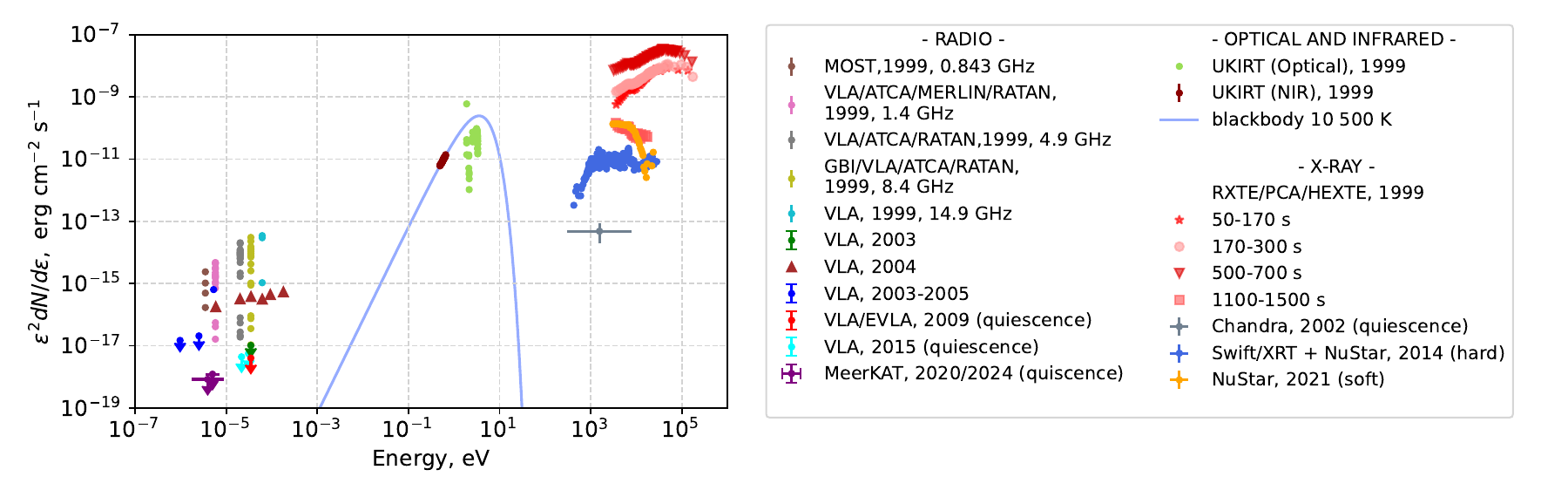}
    \caption{Combined broad-band measurements of the V4641 X-ray binary. The sources are specified in the text.}
    \label{fig:obs-spectrum}
    
    \includegraphics[width=.95\linewidth]{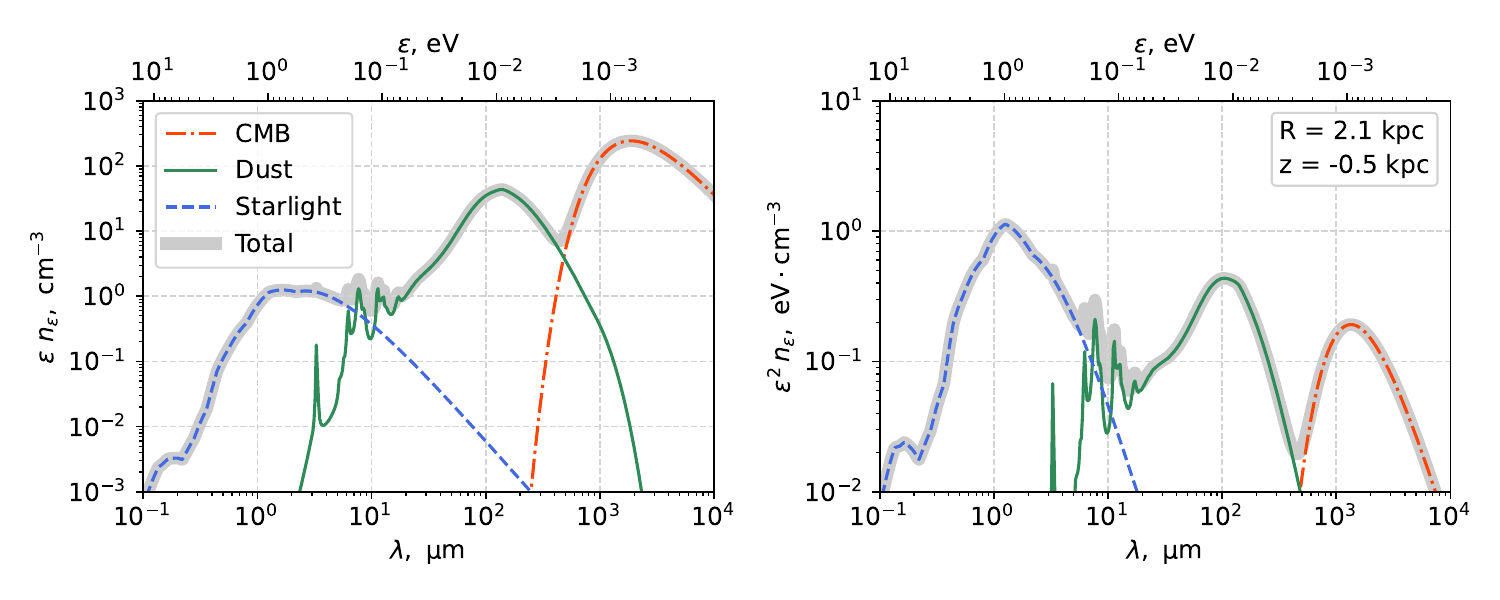}
    \caption{Background radiation field number density (left) and energy density (right) with its constituents: CMB (red dash-dotted), Dust (green, solid)~\cite{Popescu2017}, and Starlight (blue, dashed)~\cite{Popescu2017}.}
    \label{fig:bg-rad}
\end{figure*}

Angular sizes of the H.E.S.S., HAWC, and LHAASO measurements are different and increase with energy. In \autoref{fig:size-on-energy}, we show the energy dependence of the corresponding angular sizes.

In~\autoref{fig:size-on-energy}, we estimate the distances particles have travelled from the source as follows: for H.E.S.S. and HAWC, we use the quoted extensions of the single-component models of H.E.S.S. and HAWC respectively (as given in Fig.~5 of Ref.~\cite{HESS-AA2025}); for LHAASO~\cite{LHAASO2024}, we show two scales: semi-transparent points show the extensions of the best-fit single-component elliptic models in the energy bins $25\dm{TeV} < E < 250\dm{TeV}$ and $250\dm{TeV} < E < 800\dm{TeV}$ (see Sup.~Table~5 of~\cite{LHAASO2024}); opaque points correspond to the distances of the Southern edges of those ellipses from V4641~Sgr (as determined by us from Sup. Table~5 of~\cite{LHAASO2024}).

In the same figure, we also show best-fit diffusion scaling laws for Kolmogorov and Kraichnan turbulence spectra: $D\propto E^{1/3}$ and $D\propto E^{1/2}$ respectively~\cite{SemikozRev2019}. 

    \section{Luminosity in leptonic scenarios}
\label{appendix:leptonic-luminosity}

\begin{figure}[h!]
    \centering
    \includegraphics[width=\linewidth]{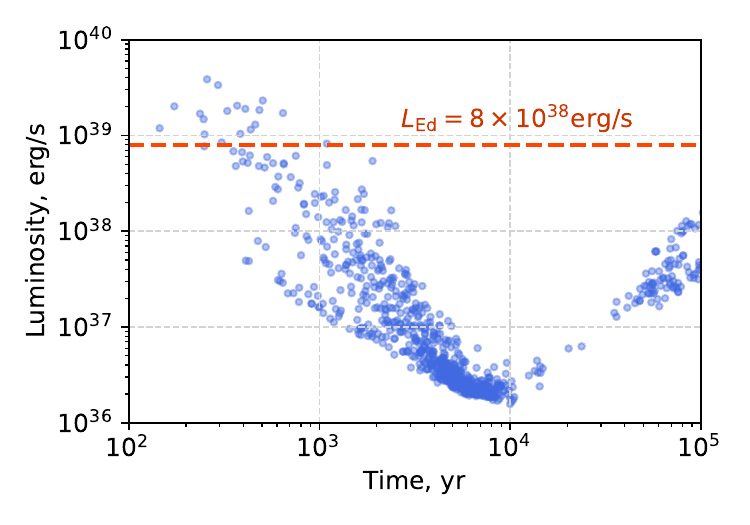}
    \vspace{-2ex}
    \caption{Distribution of sampled points on the lifetime-luminosity diagram. Eddington luminosity is marked with a dashed red line.}
    \label{fig:luminosity_on_time}
\end{figure}

As naturally expected, the increase of nebula's lifetime leads to a decrease in required luminosity. However, this simple trend changes due to intensive synchrotron cooling at high electron energies and Inverse Compton cooling, which is approximately of the same order in the energy range of our interest. MCMC modeling showed that for longer lifetimes, required luminosity again increases. This dependence is shown in \autoref{fig:luminosity_on_time} for $B = 3\dm{\mu G}.$ Under these assumptions, the least energetically demanding scenario corresponds to $t\sim 7\dm{kyr}$ and $L\sim2\times10^{36}\dm{erg/s}\sim2\times10^{-3}L_\rm{Ed}.$

    \section{1D Profiles}
\label{appendix:1dprofiles}

To calculate $1D$ profiles, we derived 1D solutions of the classical anisotropic diffusion equation~\eqref{eq:dif_eq}\vspace{-1ex}
\begin{equation}
    (\partial_t - \vec{\partial}\hat{D}\vec{\partial})n(r, t) = q(r,t),
    \label{eq:dif_eq}
    \vspace{-1ex}
\end{equation}
where $\hat{D} = \rm{diag}(D_\parallel,D_\bot,D_\bot)$ for two different extreme cases: flash-like emission \mbox{$q_0^\rm{flash}(t,\mathbf{r}) = N_0\,\delta(t)\,\delta(\mathbf{r})$}, and continuous emission \mbox{$q_0^\rm{cont}(t,\mathbf{r}) = Q_0\,\theta(t)\,\delta(\mathbf{r})$}.

Setting $z$ along the major axis of the nebula, we can introduce $x' = \sqrt{D_\perp/D_\parallel}\,x$, $y' = \sqrt{D_\perp/D_\parallel}\,y$, and $z'=z$. In these coordinates, \eqref{eq:dif_eq} becomes homogeneous diffusion equation with well-known Green's function. In these new coordinates, with $r'^2 = x'^2 + y'^2 + z'^2$, the solutions for the two aforementioned scenarios are respectively\vspace{-1ex}
\begin{gather}
    n^\rm{f}(r',t) = \frac{Qt}{(4\pi D t)^{3/2}}\exp\left\{-\dfrac{r'^2}{4Dt}\right\},\label{eq:flash_3D}\\
    n^\rm{c}(r',t) = \frac{Q}{4\pi D r'}\rm{erfc}\left(\dfrac{r'}{\sqrt{4Dt}}\right).\label{eq:continuous_3D}
\end{gather}

To obtain a 1D profile, we integrated these densities along $x'$ and $y'$: $I\propto \int dx'\,dy'\,n(r',t)$, which gives\vspace{-1ex}
\begin{gather}
    I^\rm{f} = \frac{N_0}{\sqrt{\pi D t}}\exp\left\{-\dfrac{z^2}{4Dt}\right\}\\
    I^\rm{c} = \frac{Qt}{\sqrt{4\pi D t}}\left[e^{-z^2/4Dt} - \frac{\pi|z|}{\sqrt{4\pi Dt}}\,\rm{erfc}\left(\dfrac{|z|}{\sqrt{4Dt}}\right)\right]
\end{gather}

    \section{Telegrapher's equation Green's function}
\label{appendix:telegraphers-greens-function}

To mimic the ballistic-diffusion transition in leptonic case, we used the one-dimensional telegrapher's equation \eqref{eq:1D-telegraph}, which reduces to a two-directional free-streaming with velocity $v_0$ for $t\ll\tau$ \eqref{eq:1D-free-streaming}, and to a one-dimensional diffusion equation with diffusion coefficient $D = \tau v^2$ for $t\gg\tau$ \eqref{eq:1D-diffusion}.
\begin{gather}
    \left(\partial^2_{tt} + \tau^{-1} \partial_t - v^2\partial^2_{zz}\right)n(z,t) = \left(\partial_t + \tau^{-1}\right)q(z,t),
    \label{eq:1D-telegraph}\\
    \left(\partial^2_{tt} - v^2\partial^2_{zz}\right)n(z,t) = \partial_t q(z,t)
    \label{eq:1D-free-streaming}\\
    \left(\partial_{t} - \tau v^2\partial^2_{zz}\right)n(z,t) = q(z,t)
    \label{eq:1D-diffusion}
\end{gather}

To arrive at free-streaming equation, one writes two independent transport equations for two particle populations: the ones moving along $z$-axis, denoted $n_+$, and the ones moving in the opposite direction, denoted $n_-$. By changing the variables into total concentration $n = n_+ + n_-$ and particle flux $j = (n_+ - n_-) v_0$, one recovers \eqref{eq:1D-free-streaming}.

We calculated the Green's function $G(z,t)$ for the telegrapher's equation with trivial boundary conditions, such that \mbox{$n(z,t) = \intop_{t_0}^{t}dt'\intop_{z_0}^{z}dz'\,G(z-z',t-t')q(z',t')$} solves \eqref{eq:1D-telegraph}. Using the standard Green’s function for the telegrapher’s equation (see eqs.~(7.4.18)–(7.4.27) in \cite{MorseFeshbach1953}), we obtained
\begin{multline}
    G(z,t) = \dfrac{e^{-t/2\tau}}{4v\tau}\bigg[2v\tau\delta(vt-|z|)
    + I_0\left(\xi(z,t)\,\frac{t}{2\tau}\right)+\\
    +\frac{1}{\xi(z,t)}I_1\left(\xi(z,t)\,\frac{t}{2\tau}\right)\bigg]\theta(vt-|z|),
    \label{eq:telegrapher's_green's}
\end{multline}
where $\delta(\cdot)$ is the Dirac delta function, representing the free-streaming component; $\theta(\cdot)$ is the Heaviside step function, which restricts the propagation velocity to be less than $v_0$; $I_\nu(\cdot)$, with $\nu = 0,1$, denotes the modified Bessel functions; and for compactness we introduced the dimensionless parameter \mbox{$\xi = \sqrt{1 - (z/vt)^2}$}.

In the aforementioned limiting cases, \eqref{eq:telegrapher's_green's} reduces into standard 1D free-streaming and diffusion Green's functions \eqref{eq:free-greens'} and \eqref{eq:dif-greens'} respectively.
\begin{gather}
    G_\rm{free}(z,t;v) = \dfrac{\delta(vt-|z|)}{2},
    \label{eq:free-greens'}\\
    G_\rm{dif}(z,t;D) = \dfrac{e^{-z^2/4Dt}}{\sqrt{4\pi D t}}.
    \label{eq:dif-greens'}
\end{gather}

\begin{figure}[h!]
    \centering
    \includegraphics[width=\linewidth]{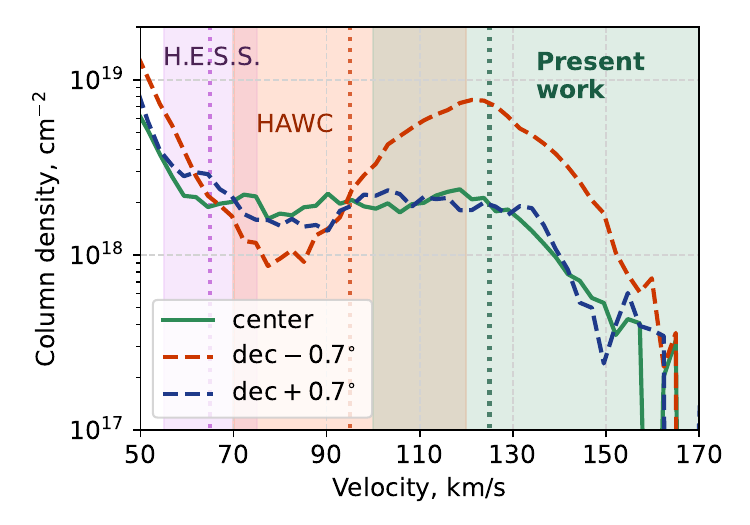}
    \vspace{-3ex}
    \caption{
    Column density on velocity at five points along the right ascension axis. The velocity ranges considered by HAWC~\cite{HAWC2024} and H.E.S.S.~\cite{HESS-AA2025} are marked in red and violet respectively. The range used in the present work is shown in green, the reference velocity of $125\dm{km/s}$ used to draw the maps on~\autoref{fig:map_comparison} is shown with a vertical green dotted line.
    }
    \label{fig:col_density_on_vel}
\end{figure}

To draw a 3D picture, as shown in the right column of \autoref{fig:two_scenarios}, we used an artificial Green's function \mbox{$\tilde{G}(x,y,z,t) = G_\rm{dif}(x,t;D_\perp)\times G_\rm{dif}(y,t;D_\perp) \times G(z,t)$} with $D_\parallel = v^2\tau$.
\\
    \begin{minipage}{\linewidth}

    \section{Cold proton density from HI4PI}
    \label{appendix:cold-proton-density}

    With the use of the HI4PI maps, we investigated cold proton density at the region of the source. To account for asymmetry, in \autoref{fig:col_density_on_vel} we present five different column density on velocity curves at the center and at $\pm0.7^\circ$ from the source along the constant right ascension line. One can notice a relative increase of the column density at the southern point in the velocity range $100 < v < 150\dm{km/s}.$
\end{minipage}

\end{document}